# Coevolving Cellular Automata with Memory for Chemical Computing: Boolean Logic Gates in the B-Z Reaction

Christopher Stone, Rita Toth, Ben de Lacy Costello, Larry Bull & Andrew Adamatzky

Unconventional Computing Group
University of the West of England
Bristol BS16 1QY, U.K.

{Christopher3.Stone, Rita.Toth, Ben.Delacycostello, Larry.Bull, Andrew.Adamatzky}@uwe.ac.uk

**Abstract.** We propose that the behaviour of non-linear media can be controlled automatically through coevolutionary systems. By extension, forms of unconventional computing, i.e., massively parallel non-linear computers, can be realised by such an approach. In this study a light-sensitive sub-excitable Belousov-Zhabotinsky reaction is controlled using various heterogeneous cellular automata. A checkerboard image comprising of varying light intensity cells is projected onto the surface of a catalyst-loaded gel resulting in rich spatio-temporal chemical wave behaviour. The coevolved cellular automata are shown to be able to control chemical activity through dynamic control of the light intensity. The approach is demonstrated through the creation of a number of simple Boolean logic gates.

## 1. Introduction

There is growing interest in research into the development of 'non-linear computers'. The aim is to harness the as yet only partially understood intricate dynamics of non-linear media to perform complex 'computations' more effectively than with traditional architectures and to further the understanding of how such systems function. Previous theoretical and experimental studies have shown that reaction-diffusion chemical systems are capable of information processing. Experimental prototypes of reaction-diffusion processors have been used to solve a wide range of computational problems, including image processing, path planning, robot navigation, computational geometry and counting (see [Adamatzky et al., 2005] for an overview). In addition to these applications, Boolean logic gates have been constructed in such excitable chemical systems (e.g., [de Lacy Costello & Adamatzky 2005]) and in bistable systems [Rössler, 1974].

In this paper, we produce networks of non-linear media — reaction-diffusion systems — to achieve user-defined computation in a way that allows direct control of the media. We use a spatially-distributed light-sensitive form of the Belousov-Zhabotinsky (BZ) [Zhaikin & Zhabotinsky, 1970] reaction which supports travelling reaction-diffusion waves and patterns. Exploiting the photoinhibitory property of the reaction, the chemical activity (amount of excitation on the gel) can be controlled by the applied light intensity, namely it can be decreased by illuminating the gel with high light intensity and vice versa. In this way a BZ network is created via light and controlled using cooperative coevolutionary computing to design heterogeneous Cellular Automata (CA) [von Neumann, 1966]. We adapt the chemical system described by Wang et al. [1999] and explore its computational potential based on the movement and control of wave fragments. In our experiments a heterogeneous CA controls the light intensity in the cells of a checkerboard image projected onto the surface of the light sensitive catalyst-loaded gel. Initially a certain number of wave fragments are created on the gel and the coevolved CA is shown able to create a number of two-input Boolean logic gates - AND, NAND and XOR - through dynamic control of the light intensity within each cell in a simulated chemical system.

Previously, several results from the evolution of CAs to perform defined tasks have been presented. Mitchell et al. (e.g., [1993][1994]) have investigated the use of a Genetic Algorithm (GA) [Holland, 1975] to learn the rules of uniform one-dimensional, binary CAs. The GA produces the entries in the update table used by each cell, candidate solutions being evaluated with regard to their degree of success for the given task — density and synchronization. Andre et al. [1999] repeated Mitchell et al.'s work, using Genetic Programming [Koza, 1992] to evolve update rules. They report similar results. Sipper (e.g., [1997]) presented a non-uniform, or heterogeneous, approach to evolving CAs. Each cell of a one- or two-dimensional CA is also viewed as a GA population member, mating only with its lattice neighbours and receiving an individual fitness. He shows an increase in performance over Mitchell et al.'s work by exploiting the potential for spatial heterogeneity in the tasks. In this paper we extend our recently presented version of Sipper's approach to control the behaviour of the B-Z system described [Stone et al., 2007].

## 2. Cooperative Coevolution of Heterogeneous CAs

The characteristics of the chosen chemical system are very much akin to those of two-dimensional cellular automata, such as the Game of Life [Gardner, 1970]. That is, fragments of excitation travel across the surface of the gel, often colliding to form other fragments or self-extinguishing, as do the gliders in "Life." Further, the light projections which cause such behaviour can be arranged in a regular grid of cells over the gel surface. We are therefore interested in using cellular automata to control the behaviour of the fragments to implement computation, particularly forms of collision-based computing (e.g., [Adamatzky, 2002]).

As noted, we have previously presented an approach to the use of a heterogeneous CA to control the BZ chemical system [Stone et al., 2007]. The heterogeneous network has a CA topology, i.e., simple finite automata are arranged in a two-dimensional lattice, with aperiodic boundary conditions (an edge cell has five neighbours, a corner cell has three neighbours, all other cells have eight neighbours each). Each automaton updates its state depending upon its own state and the states of its neighbours. States are updated in parallel and in discrete time. In this work, the transition function of every automaton cell is evolved by a simple evolutionary algorithm (EA).

This approach is very similar to that presented by Sipper [1997]. However, his reliance upon each cell having access to its own fitness means it is not applicable in the majority of chemical computing scenarios we envisage. Instead, fitness is based on emergent global phenomena in our approach (as in [Mitchell et al., 1993], for example). Thus, following Kauffman [1993], we use a simple coevolutionary approach wherein each automaton of the two-dimensional CA controller is developed via a simple genetics-based hillclimber. Due to the use of a single global fitness measure, automata do not evolve in isolation and fitness is influenced by the state of all automata cells in the grid. Hence the automata must coevolve cooperatively to solve the global task.

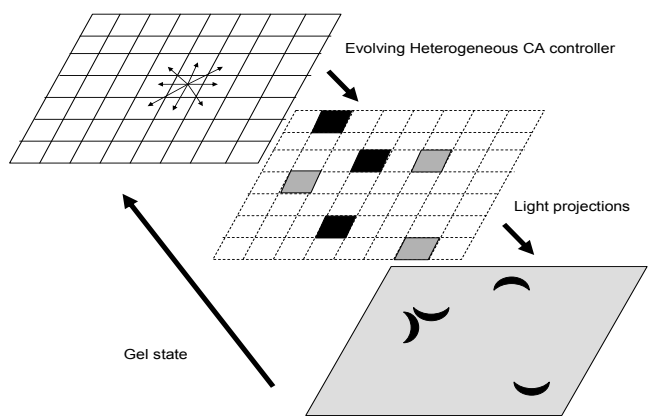

**Fig. 1**: Relationship between the CA controller, applied grid pattern and chemical system comprising one process control cycle.

For a given experiment, a random set of CA rules is created for a two-dimensional array of size 10-by-10, i.e., 100 automata, each responsible for the corresponding area of the gel surface (Figure 1). The transition state for each possible rule for an automaton is represented by a gene in the genome, which takes one of the three discrete light intensity values used in the experiment. As previously mentioned, the grid edges are not connected (i.e., the grid is planar and does not form a toroid) and the neighbourhood size of each cell is of radius 1; cells consider neighbourhoods of varying size depending upon their spatial position. The state of the gel is described by a binary string which indicates the thresholded level of chemical activity in each neighbourhood location. Previously, we showed this system capable of increasing or decreasing the amount of activity across the surface of the gel in both numerical simulation and the actual chemical system [Stone et al., 2007].

## 3. Chemical Model

Features of the chemical system are simulated using a two-variable Oregonator model modified to account for photochemistry [Field & Noyes, 1973; Krug et al., 1990; Kádár et al., 1997]:

$$\frac{\partial u}{\partial t} = \frac{1}{\varepsilon}\left(u - u^2 - (fv + \Phi)\frac{u-q}{u+q}\right) + D_u \nabla^2 u$$

$$\frac{\partial v}{\partial t} = u - v$$

The variables $u$ and $v$ represent the instantaneous local concentrations of the bromous acid autocatalyst and the oxidized form of the catalyst, $HBrO_2$ and tris (bipyridyl) Ru (III), respectively, scaled to dimensionless quantities. The rate of the photo-induced bromide production is designated by $\Phi$, which also denotes the excitability of the system. Low simulated light intensities facilitate excitation while high intensities result in the production of bromide that inhibits the process. The system was integrated using the Euler method with a five-node Laplacian operator, time step $\Delta t$=0.001 and grid point spacing $\Delta x$=0.62. The diffusion coefficient, $D_u$, of species $u$ was unity, while that of species $v$ was set to zero as the catalyst is immobilized in the gel. The kinetic parameters were set to $\varepsilon = 0.11$, $f = 1.1$ and $q = 0.0002$. The medium is oscillatory in the dark which made it possible to initiate waves in a cell by setting its simulated light intensity to zero. At different $\Phi$ values the medium is excitable, subexcitable or non-excitable. The gel surface area is represented by 200-by-200 simulation points.

## 4. Control process

Waves were initiated by setting the excitability to zero for a small area under and just outside the bottom centre of the grid. These waves were channelled into the grid and broken up into 12 fragments by choosing an appropriate light pattern as shown in Figure 2(a). The black area represents the excitable medium whilst the white area is non-excitable. After initiation three light levels were used: one is sufficiently high to inhibit the reaction; one is at the sub-excitable threshold such that excitation just manages to propagate; and the other low enough to fully enable it. The modelled chemical system was run for 600 iterations of the simulator. This value was chosen to produce network dynamics similar to those obtained in experiment over 10 seconds of real time.

A colour image was produced by mapping the level of oxidized catalyst at each simulation point into an RGB value. Image processing of the colour image was necessary to determine chemical activity. This was done by differencing successive images on a pixel by pixel basis to create a black and white thresholded image. Each pixel in the black and white image was set to white (corresponding to excitation) if the intensity of the red or blue channels in successive colour images differed by more than 5 out of 256 pixels (1.95%). Pixels at locations not meeting this criterion were set to black. An outline of the grid was superimposed on the black and white images to aid visual analysis of the results.

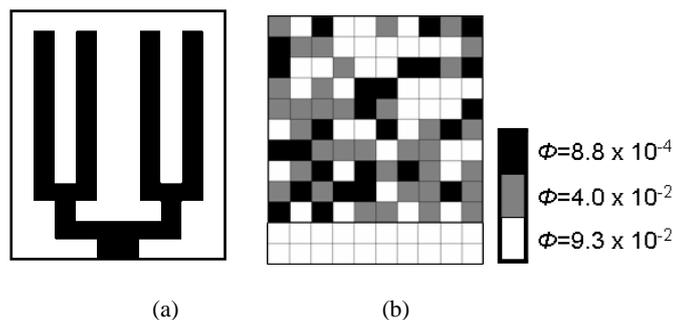

(a)          (b)
**Fig. 2:** Showing initiation pattern (a) and a typical example of a coevolved light pattern (b).

The black and white images were then processed to produce a 100-bit description of the grid for the CA. In this description each bit corresponds to a cell and it is set to true if the average level of activity within the given cell is greater than a pre-determined threshold of 10%. Here, activity is computed for each cell as the fraction of white pixels in that cell. This binary description represents a high-level depiction of activity in the BZ network and is used as input to the CA. Once cycle of the CA is performed whereby each cell of the CA considers its own state and that of its neighbours (obtained from the binary state description) to determine the light level to be used for that grid cell in the next time step. Each grid cell may be illuminated with one of three possible light levels. The CA returns a 100-digit trinary action string, each digit of which indicates whether high ($\Phi=0.093023$), sub-excitable threshold ($\Phi=0.04$) or low ($\Phi=0.000876$) intensity light should be projected onto the given cell. The progression of the simulated chemical system, image analysis of its state and operation of the CA to determine the set of new light levels comprises one control cycle of the process. A typical light pattern generated by the CA controller is shown in Figure 2(b).

Another 600 iterations are then simulated with those light-levels projected, etc. until 25 control cycles have passed. After 25 control cycles, the fitness of the emergent behaviour is calculated. As previously mentioned, the EA used in this work employs a single global fitness measure. The nature of the tasks undertaken means that it is not possible to decompose solutions obtained by the EA and apportion fitness to their constituent parts. Instead, a global fitness is determined according to how well the task has been performed and this fitness is assigned to the genome for each CA cell.

|     | $I_1$  $I_2$ |   | O (AND) |   | O (NAND) |
|-----|--------------|---|---------|---|----------|
| 00  |              | 0 |         | 1 |          |
| 01  |              | 0 |         | 1 |          |
| 10  |              | 0 |         | 1 |          |
| 11  |              | 1 |         | 0 |          |

**Fig. 3**: Typical examples of solutions of AND and NAND logic gates after 25 cycles. Input states $I_1$, $I_2$ for the logic gates are shown on the left and consist of two binary digits, spatially encoded using left and right "initiation trees". The EA found the AND solution in 56 input presentations and the NAND in 364 input presentations.

The EA is a simple hillclimber. After fitness has been assigned, some proportion of the CA's genes are randomly chosen and mutated. Mutation is the only variation operator used here to modify a given CA cell's transition rule to allow the exploration of alternative light levels for the grid state. For a CA cell with eight neighbours there are $2^9$ possible grid state to light level transitions, each of which is a potential mutation site. After the defined number of such mutations have occurred, a generation of the EA is complete and the simulation is reset and repeated as described.

The EA keeps track of which CA states are visited since mutation. On the next fitness evaluation (at the end of a further 25 control cycles) mutations in states that were not visited are discarded on the grounds that they have not contributed to the global fitness value and are thus untested. We also performed control experiments with a modified version of the EA to determine the performance of an equivalent random CA controller. This algorithm ignored the fitness of mutants and retained all mutations except those from unvisited states.

## 5. Chemical Logic Gates

We have designed a simple scheme to simulate a number of two-input Boolean logic gates under the framework described above where excitation is fed in at the bottom of the grid into the branching pattern. To encode a logical '1' and '0' either both branches or just one branch of the two "trees" shown in Figure 2(a) are allowed to fill with excitation, i.e., the grid is divided into two for the inputs (Figure 3). The number of active cells in the whole grid, that is those with activity at or above the 10% threshold, is used to distinguish between a logical '0 and '1' as the output of the system. For example, in the case of XOR, the CA controller must learn to keep the number of active cells below the specified level for the 00 and 11 cases but increase the number for the 01 and 10 case.

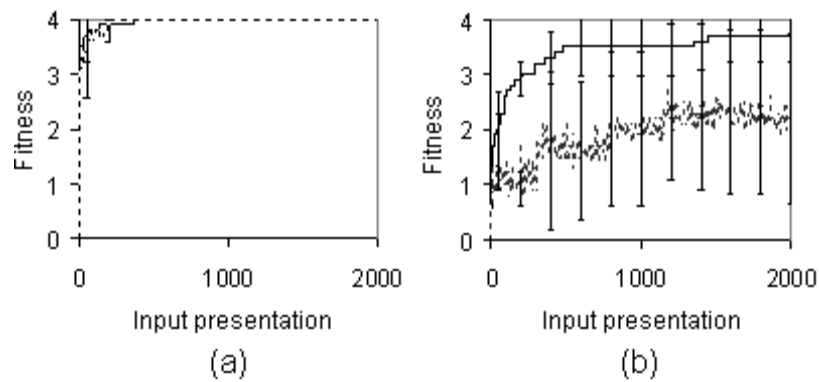

**Fig. 4:** Showing the average fitness over time for 10 runs for the (a) AND gate and (b) NAND gate on the simulated chemical system. Dashed lines: random controller (10 runs).

Figure 3 shows typical examples of the logic gates learned using the simulated chemical system. Here the mutation rate was set at 4000 genes per EA generation. The required number of active cells was set at 20. Each of the four possible input combinations is presented in turn – 00 to 11 – and for each input presentation the system is allowed to develop for 25 control cycles. Fitness of the logic gate is evaluated after the complete sequence of four input presentations. Each correct output scores 1, resulting in a maximum possible fitness of 4 for a correctly functioning gate. Figure 4(a) and (b) show the fitness averaged over ten runs for the AND and NAND tasks, with similar results found for XOR (not shown).

**Table1**. Performance on AND, NAND and XOR gates.

| Gate | Controller | Success rate | Min. | Max. | Avg. | Std. |
|------|------------|--------------|------|------|------|------|
| AND  | Coevolutionary | 10/10 | 8 | 144 | 61 | 45.69 |
|      | Random | 10/10 | 4 | 200 | 64 | 71.73 |
|      | Simple Memory | 10/10 | 32 | 252 | 140 | 78.88 |
|      | WH Memory | 10/10 | 4 | 68 | 34 | 21.35 |
| NAND | Coevolutionary | 7/10 | 288 | >2000 | 1065 | 767.21 |
|      | Random | 4/10 | 300 | >2000 | 1454 | 744.49 |
|      | Simple Memory | 2/10 | 632 | >2000 | 1858 | 431.08 |
|      | WH Memory | 10/10 | 16 | 720 | 228 | 234 |
| XOR  | Coevolutionary | 9/10 | 348 | >2000 | 808 | 510.08 |
|      | Random | 10/10 | 20 | 1080 | 455 | 333.68 |
|      | Simple Memory | 8/10 | 48 | >2000 | 1326 | 784.74 |
|      | WH Memory | 10/10 | 8 | 428 | 148 | 131.16 |

Table 1 shows a more detailed comparison, namely the results of ten runs for each gate. Due to the high computational requirements needed to perform the simulations a limited number of input presentations were allowed for each experiment and an experiment was considered successful if the controller found a solution within 2000 input presentations. The success rate shows the number of successful runs out of ten. The AND task was so simple that a solution was easily found even with a random controller. This is because the first three inputs provided activity levels similar to the correct outputs, and only the activity levels provided by the 11 input needed to be changed to generate appropriate output activity, namely the controller had to increase excitation to get higher than the required number of active cells (that is, those with an activity level greater than or equal to 10%). In contrast, the NAND gate was the most difficult task, because the controller had to achieve the opposite activity levels to those provided by the input states. For 00, 01 and 10 inputs the initial number of fragments were less than the required value so the controller had to increase the excitation to get the correct logical '1' output, while for the 11 input the controller had to decrease the excitation to achieve the logical '0' output. These results indicate the ability of the coevolutionary approach for universal computation since all functions can be constructed by NAND gates. The XOR task was also hard because the activity levels provided by three of the initial inputs (00, 01, and 10) were the opposite of the desired output activity levels and only the 11 input provided an appropriate direct basis for correct output activity.

In the cases where the success rate was less than ten, the averages in Table 1 are the lowest possible averages, since 2000 was taken as the number of input presentations required, even though no solution was found in these cases. For this reason we can only use these data as an indication of the difficulty of the task.

## 6. Coevolving CAs with Memory

As discussed above, the BZ reaction exhibits rich spatio-temporal behaviour. Recently, the standard CA framework has been extended to explicitly consider temporal dynamics in the transition rule by the inclusion of memory mechanisms (e.g., [Alonso-Sanz, 2004]). Given the strong temporal element of BZ systems, we have explored the utility of including memory within the evolving heterogeneous CA controller.

A simple way of implementing memory is for the CA transition rule $\Phi$ to consider the neighbourhood $N$ of a cell $i$ supplemented with the state $\sigma$ of the cell on the previous cycle:

$$\sigma_i^{t+1} = \Phi(\sigma_j^t \in N_i, \sigma_i^{t-1})$$

However, this means that the size of the CA's genome must increase to incorporate the extra state and the size of the search space is doubled. To overcome this limitation we have also implemented a form of memory using the well-known Widrow-Hoff Delta rule with learning rate $\beta=0.2$. This provides a weighted average memory with no representational overhead in the genome. For this type of memory:

$$m_i^0 = 0.5$$

$$m_i^{t+1} = m_i^t + \beta(\sigma_i^t - m_i^t)$$

$$s_i^t = \begin{cases} 1 & \text{if } m_i^t > 0.5 \\ 0 & \text{if } m_i^t \leq 0.5 \end{cases}$$

$$\sigma_i^{t+1} = \Phi(s_j^t \in N_i)$$

As Table 1 shows, the simple explicit memory scheme degrades performance but the weighted average scheme improves performance in all cases. Moreover, t-test results with $\alpha = 0.01$ for the NAND and XOR gates show a statistically significant performance improvement when using the Widrow-Hoff memory scheme. Thus, it would appear that the inclusion of memory can enable the CA to capture better the temporal dynamics of the reaction, as envisaged. However, it is apparent from these initial results that factors such as the type and/or depth of memory are important in achieving a benefit.

## 7. Conclusions

Excitable and oscillating chemical systems have previously been used to solve a number of simple computational tasks. However the experimental design of such systems has typically been non-trivial. In this paper we have presented results from a methodology by which to achieve the complex task of designing such systems — through the use of coevolution. We have shown using a simulated system that it is possible to control the behaviour of a light-sensitive BZ reaction to implement a number of Boolean logic gates. We have also shown that the inclusion of memory within such discrete dynamical systems can better enable them to control such non-linear media. Current work is utilising the actual chemical system and exploring the utility of memory mechanisms within CAs to control and model complex systems in general.

## Acknowledgements

This work was supported by EPSRC Grants no.'s EP/E049281/1 and GR/T11029/1.

## References

Adamatzky, A. (2002) (ed.) *Collision-based Computing*. Springer.
Adamatzky, A., De Lacy Costello, B. & Asai, Y. (2005) *Reaction Diffusion Computers*. Elsevier.
Alonso-Sanz, R. (2004) One-dimensional, $r$=2 cellular automata with memory. *International Journal of Bifurcation and Chaos* 14: 3217-3248.
Andre, D., Koza, J.R., Bennett, F.H. & Keane, M. (1999) *Genetic Programming III*. MIT Press.
De Lacy Costello, B. & Adamatzky, A. (2005) Experimental implementation of collision-based gates in Belousov–Zhabotinsky medium. *Chaos Solitons and Fractals* 25: 535-544.
Field, R. J. & Noyes, R. M. (1973) Oscillations in chemical systems. IV. Limit cycle behavior in a model of a real chemical reaction. *J. Chem. Phys*. 60: 1877-1884.
Gardner, M. (1970) The Fantastic Combinations of John Conway's New Solitaire Game 'Life' *Scientific American* 223(4): 120-123
Holland, J.H. (1975) *Adaptation in Natural and Artificial Systems*. University of Michigan Press.
Kauffman, S. A. (1993) *The Origins of Order: Self-Organization and Selection in Evolution*. Oxford.
Koza, J.R. (1992) *Genetic Programming*. MIT Press.
Krug, H.-J., Pohlmann, L. & Kuhnert, L. (1990) Analysis of the modified complete Oregonator accounting for oxygen sensitivity and photosensitivity of Belousov-Zhabotinsky systems*, J. Phys. Chem*. 94: 4862-4866.
Mitchell,M., Hraber P, & Crutchfield J. (1993) Revisiting the Edge of Chaos: Evolving Cellular Automata to Perform Computations. *Complex Systems* 7: 83-130.
Mitchell, M., Crutchfield, J. & Hraber, P. (1994) Evolving Cellular Automata to Perform Computations: Mechanisms and Impediments. *Physica D* 75: 361-391.
Rössler, O. (1974) In M. Conrad, W. Güttinger, & M. Dal Cin (eds) *Physics and Mathematics of the Nervous System*. Springer.
Sipper, M. (1997) *Evolution of Parallel Cellular Machines*. Springer.
Stone, C., Toth, R., Adamatzky, A., Bull, L. & De Lacy Costello, B. (2007) Towards the Coevolution of Cellular Automata Controllers for Chemical Computing with the B-Z Reaction. In D. Thierens et al. (eds) *GECOO-2007: Proceedings of the Genetic and Evolutionary Computation Conference*. ACM Press, pp472-478
Von Neumann, J. (1966) *The Theory of Self-Reproducing Automata*. University of Illinois.
Wang, J., Kádár, S., Jung, P. & Showalter, K. (1999) Noise driven Avalanche behavior in subexcitable media. *Physical Review Letters* 82: 855-858.
Zaikin, A. N. & Zhabotinsky, A. M. (1970) Concentration wave propagation in two-dimensional liquid-phase self-oscillating system. *Nature* 225: 535-537.